\documentstyle[aps,prl,twocolumn]{revtex}

\newcommand{\ket}[1]{|#1\rangle}

\newcommand{\kettwo}[2]{|#1#2\rangle}

\newcommand{\ketthree}[3]{|#1#2#3\rangle}

\newcommand{\be}{\begin{equation}}
\newcommand{\ee}{\end{equation}}
\newcommand{\bea}{\begin{eqnarray}}
\newcommand{\eea}{\end{eqnarray}}

\begin{document}


\draft

\title{Generalized Schmidt decomposition and classification of three-quantum-bit states}

\author{A. Ac\'{\i}n$^\dagger$, A. Andrianov$^{\dagger\ast}$, L. Costa$^{\ddagger}$, E. Jan\'e$^{\dagger}$, J.I. Latorre$^{\dagger}$ and\,R. Tarrach$^{\dagger}$}

\address{$^\dagger$Departament d'Estructura i Constituents de la
Mat\`eria, Universitat de Barcelona, Diagonal 647, E-08028 Barcelona, Spain.\\
$^\ddagger$Departament d'\`Algebra i Geometria, Universitat de Barcelona,Gran Via Corts Catalanes 585, E-08007 Barcelona, Spain.\\
$^\ast$Department of Theoretical Physics, St. Petersburg State University, 198904, St. Petersburg, Russia. \\
e-mail: ejane@ecm.ub.es
}

\date{\today}

\maketitle


\begin{abstract}
We prove for any pure three-quantum-bit state the existence of 
local bases which allow one to build a set of five orthogonal 
product states in terms of which the state can be written 
in a unique form. This leads 
to a canonical form which generalizes the two-quantum-bit Schmidt 
decomposition. It is uniquely characterized by the five entanglement 
parameters. It leads to a complete classification of the three-quantum-bit 
states. It shows that the right outcome of an adequate local 
measurement always erases all entanglement between the other 
two parties.
\end{abstract}


\bigskip 



The Schmidt decomposition \cite{schmidt,peres} allows one to write any pure
state of a bipartite system as a linear combination of biorthogonal 
 product states or, equivalently, of a
non-superfluous set of product states built from local bases. For
two quantum-bits (qubits) it reads
\be
\ket\Psi = \cos\theta\, \kettwo{0}{0}  + \sin\theta\, \kettwo{1}{1}
\hspace{4pt}  ,  \hspace{4pt} 0\le\theta\le\pi/4.
\label{one}
\ee
Here $\kettwo{i}{i}\equiv\ket{i}_A{\otimes}\ket{i}_B$, both local
bases $\{\ket{i}\}_{A,B}$ depend on the state $\ket\Psi$, the
relative phase has been absorbed into any of the local bases, and
the state $\kettwo{0}{0}$ has been defined by carrying the larger
(or equal) coefficient. A larger value of $\theta$ means more
entanglement. The only entanglement parameter, $\theta$, plus the
hidden relative phase, plus the two parameters which
define each of the two local bases are the six parameters of any two-qubit
pure state, once normalization and global phase have been disposed
of.

Very many results in quantum information theory have been obtained
with the help of the Schmidt decomposition: its simplicity
reflects the simplicity of bipartite systems as compared to
N-partite systems. Much of its usefulness comes from it not being
superfluous: to carry one entanglement parameter one needs only
two orthogonal product states built from local bases states, no
more, no less.
 
The aim of this work is to generalize the Schmidt decomposition of
(\ref{one}) to three qubits. It is well known \cite{peres} that its
straightforward generalization, that is, in terms of triorthogonal product
states, is not possible (see also \cite{genschm}). Nevertheless, 
having a minimal canonical form in which to cast any pure state, by performing 
local unitary transformations, will provide a new tool for
quantifying entanglement for three qubits, a notoriously difficult
problem. It will lead to a complete classification of
exceptional states which, as we will see, is much more complex
than in the two-qubit case. The generalization to $N$ quantum dits
($d$-state systems) is not completely straightforward and will be
given elsewhere.

Linden and Popescu \cite{pop} and Schlienz \cite{sch} showed that
for any pure three-qubit state the number of entanglement
parameters is five and, using repeatedly the two-qubit Schmidt
decomposition, proved the existence for any pure state of a
reference form in terms of six orthogonal product states built
from local bases. The five entanglement parameters are one phase
(all others can be absorbed) and four moduli of the six
coefficients, so that a further constraint beyond the
normalization exists. In other words, exactly as (\ref{one}) shows
that local unitary transformations allow to make two of the
four components vanish (corresponding to $\kettwo{0}{1}$ and
$\kettwo{1}{0}$) for a two-qubit pure state, Linden, Popescu and
Schlienz proved that, also for a three-qubit system two  of the,
now eight, components can be made zero. However, the set of six
states is superfluous in the sense that its coefficients require a
constraint to lead to a unique representative of any pure state.
It is not clear whether this is the best one can do, i.e. whether
the set is minimal. We will now prove that indeed, combining
adequately the local changes of bases corresponding to
$U(1)\times{SU(2)}\times{SU(2)}\times{SU(2)}$ transformations,
 one can always do with five terms, which precisely can carry only five
entanglement parameters, leading thus to a non-superfluous unique
representation.

Notice that a straightforward counting of parameters shows that 
a nonsuperfluous set will have five states, i.e. three vanishing 
coefficients. There exist three inequivalent sets of five 
local bases product states 
\bea
\label{two} &\{
\ketthree000,\ketthree001,\ketthree010,\ketthree100,\ketthree111\}&
\nonumber \\ &\{ \ketthree000 , \ketthree001 , \ketthree110 ,
\ketthree100 , \ketthree111 \}& \\
&\{\ketthree000,\ketthree100,\ketthree110,\ketthree101,\ketthree111\}&.\nonumber
\eea
Whereas the first set is symmetric under permutation of parties, the other
two are not.

The nonequivalence of the three sets follows from the different
degrees of orthogonality between the five states within each set.
One can also readily check that all three sets can carry exactly
five entanglement parameters, four moduli and one phase, and are thus
nonsuperfluous. This is of course no proof that any state can
always be written as a linear combination of the five states of
one and the same set. We will now prove that it can always be done
for the last two sets, or their versions obtained by permuting parties.


As an introduction let us first present a one-line proof of the Schmidt
decomposition of a two-qubit state, Eq. (\ref{one}). Writing any
state in a basis of product states built from any two local bases,
\be
\ket\Psi=\sum_{i,j}t_{ij}\kettwo{i}{j},
\ee
calling $T$ the matrix of elements $t_{ij}$, and recalling that for any
$T$ there always exist two unitary matrices which diagonalize it,
\be
U_1TU_2=D,
\label{four}
\ee
 the Schmidt decomposition follows at once. Note that $U_1$ and $U_2$
correspond to the local basis changes necessary for casting the original
state into its Schmidt form.


For a three-qubit state the proof goes as follows, from
\be
\label{five}
 \ket\Psi=\sum_{i,j,k}t_{ijk}\ketthree{i}{j}{k}, \ee
one introduces the matrices $T_0$ and $T_1$ with elements
\be
\label{six}
 (T_i)_{jk}\equiv t_{ijk}. \ee
Consider now the unitary transformation on the first qubit,
\be
\label{seven}
 T_i'=\sum_{j}u_{ij}T_j, 
\ee 
such that
\be
\det\,T'_0\,=0.
 \label{eight} \ee
Notice that (\ref{eight}) has always two solutions. The
matrix obtained from $T'_0$ after diagonalization following (\ref{four}),
which corresponds to unitary transformations on the last two qubits, has
at least three zeros,
\be
\label{nine}
{(D'_0)}_{01}={(D'_0)}_{10}={(D'_0)}_{11}=0.
 \ee
This finishes the proof that any pure state of three qubits can
always be written as a linear superposition of the five states of
the last set of (\ref{two}).

The generalization to three qubits of the Schmidt decomposition, i.e. one
more zero for one more qubit, thus reads
\bea
\ket\Psi&=&\lambda_0\ketthree{0}{0}{0}+\lambda_1e^{i\varphi}\ketthree{1}{0}{0}+\lambda_2\ketthree{1}{0}{1}+\lambda_3\ketthree{1}{1}{0}+\lambda_4\ketthree{1}{1}{1} \nonumber \\
\label{ten}
&&\lambda_i\geq0\hspace{2pt}\hspace{2pt},\hspace{2pt}0\leq\varphi\leq\pi\hspace{2pt},\hspace{2pt}\mu_i\equiv\lambda_i^2,\hspace{2pt}\sum_i\mu_i=1, 
\eea
where we have chosen the second coefficient to
carry the only relevant phase, whose range, to be proven later, is also
given. Notice that we have singled out party A in obtaining
(\ref{ten}), but we could have chosen any of the three parties. 

An immediate and important consequence of this decomposition is
that there always exists for any state $\ket\Psi$ and any
(genderless) party $X$ a state $\ket{0}_X$ such that
$_X\langle{0}\ket\Psi$ is a product state of the other two parties
(unless party $X$ is not entangled with the other two parties).
That is, party $X$, knowing $\ket\Psi$, can perform a
local measurement which, for one outcome, allows it to be sure that
the other two parties share no entanglement whatsoever. Note that 
when (\ref{eight}) displays two different solutions, two such states exist.
This property suggests some applications to quantum information processing. 
It also leads to an efficient algorithm for computing the
$\lambda$'s and $\varphi$.


There is one small hitch left: as (\ref{eight}) has generically two
different solutions, any state can be written in the form of (\ref{ten})
with two different sets of coefficients. Let us dispose generically of
this redundancy. Recall that after diagonalization of $T'_0$ we are left
with the matrices
\be
 M_0 \equiv D_0'= \left ( \begin{array}{cc} \lambda_0 & 0 \\
                                                    0 & 0
                                                    \end{array}
                                                    \right ),
 \quad  M_1 = \left ( \begin{array}{cc} e^{i \varphi} \lambda_1 & \lambda_2 \\
                                                    \lambda_3 & \lambda_4
                                                    \end{array}
                                                    \right ), \label{eleven} \ee
for one solution of Eq. (\ref{eight}) and
\be
 {\tilde M}_0 = \left ( \begin{array}{cc} \tilde{\lambda}_0 & 0 \\
                                                    0 & 0
                                                    \end{array}
                                                    \right ),
 \quad \tilde{M}_1 = \left ( \begin{array}{cc} e^{i \tilde{\varphi}} \tilde{\lambda}_1 & \tilde{\lambda}_2 \\
                                                    \tilde{\lambda}_3 & \tilde{\lambda}_4
                                                    \end{array}
                                                    \right ), \label{twelve} \ee
for the other solution. Of course, both solutions can be related
by a $U(1) \times SU(2) \times SU(2) \times SU(2)$ transformation:
\be
\label{thirteen}
\begin{array}{l}
 {\tilde M}_0= e^{i \omega} U_1 ( u_{00}M_0 + u_{01}M_1) U_2
 \\
 {\tilde M}_1= e^{i \omega} U_1 ( -u_{01}^{*}M_0 + u_{00}^{*}M_1) U_2,
 \end{array}
 \ee
and the inverse
 \be
 \label{fiveteen}
 \begin{array}{l}
  M_0= e^{-i \omega} U_1^{\dagger} ( u_{00}^{*} \tilde{M}_0 - u_{01}\tilde{M}_1)
  U_2^{\dagger}
 \\
   M_1= e^{-i \omega} U_1^{\dagger} ( u_{01}^{*} \tilde{M}_0 + u_{00} \tilde{M}_1) U_2^{\dagger}.
  \end{array}
   \ee
The condition $\det M_0\,=\det\tilde{M}_0\,=0$ leads to
  \be
    \label{seventeen}
    u_{00}= -\frac{\det\,M_1}{ \lambda_0\lambda_4}u_{01} \qquad
    u_{00}^{*}= \frac{\det\tilde{M_1}}{ \tilde{\lambda}_0
    \tilde{\lambda}_4}u_{01}.
    \ee
It is tedious, but straightforward, to solve the previous
    equations. Here we only need the following results
    \be
    \label{eighteen}
    \lambda_0 \lambda_4 = \tilde{\lambda}_0  \tilde{\lambda}_4,
    \qquad u_{01}^{*}=-u_{01},
        \ee
which, from  Eq. (\ref{seventeen}), imply
 \be
 \label{nineteen}
 \det\,M_1=(\det\,\tilde{M}_1\,)^{*}.
 \ee
From here it follows that
 \be
 \label{twenty}
 \begin{array}{lcl}
 0 < \varphi < \pi  & \Leftrightarrow  & \pi < \tilde{\varphi} < 2 \pi
 \\
  0 < \tilde{\varphi} < \pi  & \Leftrightarrow  & \pi < \varphi < 2 \pi,
 \end{array}
 \ee
 so that one can always choose the solution for which
  \be
 \label{twentyone}
  0 \leq  \varphi \leq \pi,
 \ee
 which explains the range of $\varphi$ given in Eq.
 (\ref{ten}).

 
Let us mention here that by performing a unitary 
transformation on the third qubit, 
\be
|0'\rangle=\frac{1}{\sqrt{\mu_1+\mu_2}}(\lambda_1e^{i\varphi}|0\rangle+\lambda_2|1\rangle),
\ee
the decomposition for the second set of (\ref{two}) is obtained.
In the remainder we will use the first decomposition (\ref{ten}), which
is physically and mathematically more convenient.


A generalization of the Schmidt decomposition is thus given by
(\ref{ten}); any state can be written in this minimal form,
 generically in a unique way. The explicit algorithm for
 constructing this canonical form follows from the set of
 Eqs. (\ref{five}-\ref{eight}). However, particular states can
 be obtained for different values of the five entanglement
 parameters. It is thus useful to have five independent invariants
 for the classification of states which we will obtain from
 (\ref{ten}). We will take here the five minimal
 polynomial invariants of \cite{sud}.

 Defining $\Delta \equiv |\lambda_1 \lambda_4
 e^{i \varphi} -\lambda_2 \lambda_3|^2$ we find
\be
\label{twentytwo}
\begin{array}{l}
\frac{1}{2} \leq I_1 \equiv Tr \rho_A ^2= 1-2 \mu_0(1-\mu_0-\mu_1)
 \leq 1 \\
\frac{1}{2} \leq I_2 \equiv Tr \rho_B ^2= 1-2
 \mu_0(1-\mu_0-\mu_1-\mu_2)-2 \Delta
 \leq 1 \\
\frac{1}{2} \leq I_3 \equiv Tr \rho_C ^2= 1-2
 \mu_0(1-\mu_0-\mu_1-\mu_3)-2 \Delta
 \leq 1 \\ 
\frac{1}{4} \leq I_4 \equiv Tr( \rho_A \otimes \rho_B \, \rho_{AB})\\
\phantom{\frac{1}{4} \leq I_4 } =1+\mu_0(\mu_2 \mu_3 -\mu_1 \mu_4 -2\mu_2-3\mu_3-3\mu_4)\\
\phantom{\frac{1}{4} \leq I_4 =} -(2-\mu_0)\Delta\leq 1\\
0 \leq I_5 \equiv |{\rm Hdet}(t_{ijk})|^2=  \mu_0^2 \mu_4^2 \leq 
 \frac{1}{16}, 
\end{array}
\ee
where 
\bea
&\rho_{AB}\equiv Tr_C|\Psi\rangle\langle\Psi| \hspace{8pt} \rho_{C}\equiv Tr_{AB}|\Psi\rangle\langle\Psi|& \nonumber\\
&\rho_{A}\equiv Tr_B\rho_{AB} \hspace{8pt} \rho_{B}\equiv Tr_A\rho_{AB}, &
\eea
and Cayley's hyperdeterminant, Hdet$(t_{ijk})$, can be found in \cite{gel} and 
corresponds to the three-tangle of \cite{sud,cof}.

Although these five invariants are computationally simple and
 physically meaningful, as they give local information, it can be
 convenient to trade them, recalling $\sum_{i} \mu_i =1$, for
 algebraically simpler ones:
 \be
 \label{twentyfour}
 \begin{array}{l}
 0 \leq J_1 \equiv \Delta \leq \frac{1}{4} \\
  0 \leq J_2 \equiv \mu_0 \mu_2 \leq \frac{1}{4} \\
  0 \leq J_3 \equiv \mu_0 \mu_3 \leq \frac{1}{4} \\
  0 \leq J_4 \equiv \mu_0 \mu_4 \leq \frac{1}{4} \\
   J_5 \equiv \mu_0(\Delta+ \mu_2 \mu_3-\mu_1 \mu_4).
 \end{array}
 \ee
The invariants $J_4$ and $J_5$ are symmetric under permutation of parties,
while $J_1(J_2,J_3)$ is symmetric under exchange of parties B and C (A and C,
A and B).

 
We can now proceed with the complete classification of
 nongeneric three-qubit states with the help of Eqs.
 (\ref{ten}) and (\ref{twentyfour}):

\medskip

\hskip .4cm\vtop{\hsize 7.4cm
\noindent{\bf Type 1  (product states):}
$J_i=0$ for $i=1,2,3,4,5$.

\medskip
\noindent {\bf Type 2a (biseparable states):}
 $J_i=0$ except $J_1 (J_2,J_3)$ when party A(B,C) is 
not entangled with the other two parties. 

\noindent They carry only bipartite
 entanglement and depend on one parameter.

\medskip
\noindent {\bf Type 2b (generalized GHZ states):} 
$J_i=0$ except $J_4$. 

\noindent They include the standard GHZ states
 \cite{gre} and depend on one parameter.

\medskip
\noindent {\bf Type 3a (tri-Bell states):}
$\mu_1=\mu_4=0$.

\noindent
 It implies $J_4=0$, $J_1J_2+J_1J_3+J_2J_3= \sqrt{J_1J_2J_3}=\frac{J_5}{2}$.
 They depend
  on two parameters.

\medskip
\noindent {\bf Type 3b (extended GHZ states):} 
$\mu_j=\mu_k=0$, for $j,k\in 
\{1,2,3\}$ and $j\neq k$.

 \noindent It implies
$J_j=J_k=J_5=0$. They depend on two parameters and correspond to the slice states of
 \cite{car}.

\medskip
\noindent
 {\bf Type 4a:} 
$\mu_4=0$.

\noindent 
It follows $J_4=0$ and
 $\sqrt{J_1J_2J_3}=\frac{J_5}{2}$. They depend on three parameters.

\medskip

\noindent
 {\bf Type 4b:}
$\mu_2=0$ $  (\mu_3=0)$.

\noindent Then, 
  $J_2=J_5=0$ $ (J_3=J_5=0)$. They depend on three parameters.

\medskip
\noindent
 {\bf Type 4c:}
$\mu_1=0$.

\noindent  Then, $J_1(J_2+J_3+J_4)+J_2J_3= \sqrt{J_1J_2J_3}=\frac{J_5}{2}$ and they depend on three parameters.

\medskip
\noindent
 {\bf Type 5 (real states):}
$\varphi=0,\pi$.

\noindent It  implies $\sqrt{J_1J_2J_3}=\frac{J_5}{2}$. They depend on four parameters and they
 are, generically, on the boundary of the state space in the space
 of the five invariants.

}
\bigskip

Notice that the type-number indicates how many of the five states
of (\ref{ten}) characterize the states of that type. 
Because of the asymmetric character of the decomposition (\ref{ten}), some of the
states included in type 5 can be written in terms of four states, had we singled 
out party B or C \cite{nos}. Notice also that, in some
sense, the $J_i$'s are indicators of entanglement: only when 
all of them vanish there is no entanglement at all, $J_1(J_2,J_3)$
indicate bipartite entanglement and $J_4$ indicates GHZ-entanglement.


Let us further exploit our previous results.
An alternative generalization of the Schmidt decomposition could be
writing the state as a superposition of two nonorthogonal product states 
which are not built from local bases, 
\be
\label{altsch}
|\Psi\rangle=\alpha\,|a\,b\,c\rangle+\beta\,|a'b'c'\rangle,
\ee
with $\alpha$ and $\beta$ real.

Beside the trivial cases of type-1 and type-2a states,  this decomposition 
is always possible except for a familly of states depending on three 
parameters \cite{vidal}. Our decomposition allows to reproduce this result
and shows that (\ref{altsch}) is not possible when $I_5$=0 (corresponding
to type-3a and type-4a states). It can be proved that when $I_5=0$ the two 
solutions of (\ref{eight}) coincide. The same happens had we chosen to
single out any of the other parties. Therefore, 
for any party $X$, there is only one state $|0\rangle_X$ such that 
$_X\langle0|\Psi\rangle$ is a product state of the other two parties.    
Since (\ref{altsch}) implies two such states, e.g. $|a_\bot\rangle_A$ and 
$|a_\bot'\rangle_A$, it follows that type-3a and type-4a states 
cannot be written as a  sum of two nonorthogonal 
product states. When the decomposition  
(\ref{altsch}) is possible, our results give  the constructive method to 
obtain it. 
From (\ref{ten}), the second coefficient can be split into two terms
\bea
\label{split}
|\Psi\rangle&=& \left(\lambda_0|000\rangle+\frac{\lambda_1\lambda_4e^{i\varphi}-\lambda_2\lambda_3}{\lambda_4}|100\rangle \right) \nonumber \\
&&+\left(\frac{\lambda_2\lambda_3}{\lambda_4}|100\rangle+\lambda_2|101\rangle+\lambda_3|110\rangle+\lambda_4|111\rangle \right).
\eea
It is easy to see that (\ref{split}) corresponds to the sum of two 
nonorthogonal product states as (\ref{altsch}) with coefficients
\bea
\label{sol}
\alpha&=&\frac{1}{\lambda_4}\sqrt{J_1+J_4}\nonumber\\
\beta&=&\frac{1}{\lambda_4}\sqrt{\mu_2\mu_3+\mu_4(\mu_4+\mu_2+\mu_3)}.
\eea
This decomposition is unique. The states that appear in (\ref{altsch})
are orthogonal to the ones that allow each party to destroy the 
entanglement between the other two parties with some non-vanishing probability.


A final consequence of (\ref{ten}) is that, by using the bipartite 
Schmidt decomposition, any pure state can be written as a superposition of 
a product state and a biseparable state, i.e.
\bea
|\Psi\rangle = \cos \theta |000\rangle + \sin\theta|1\rangle(\cos\omega
|0'0''\rangle + \sin\omega|1'1''\rangle),
\eea
which is the minimal decomposition in terms of orthogonal product states.
It exhibits explicitly two of the five entanglement parameters. The other 
three are hidden in the moduli of the scalar products $\langle0|0'\rangle$
and $\langle0|0''\rangle$, and in one phase absorbed by one of the 
local bases.  It is also a nonsuperfluous form, though not built
from local bases.


In this work we have found the minimal decomposition of any pure three-qubit
state in terms of orthogonal product states built from local
bases. It generalizes the Schmidt decomposition and
leads to a complete classification of pure three-qubit states,
which fine grains the fully inseparable states class of the
general entanglement classification of mixed three-qubit states
\cite{dur}. Our decomposition shows that any party can, performing a
 clever local measurement, kill the entanglement between the other two
parties with nonvanishing probability. A decomposition in terms 
of the  minimal number of orthogonal product states has also been found.

Finally, we have explored whether a pure three-qubit state can be written as
a sum of two nonorthogonal product states, which can be thought as 
an alternative generalization of the Schmidt decomposition. We have
verified that only a subfamily depending on three parameters cannot 
be expressed in this form \cite{vidal}, corresponding to states with $I_5=0$.


\bigskip

The authors thank Guifr\'e Vidal and Sandu Popescu for useful discussions.
J.I.L. and R.T. acknowledge financial support by CICYT project AEN 98-0431, CIRIT project 1998SGR-00026 and CEC project IST-1999-11053, A. Andrianov by RFBR 99-01-00736 and CIRIT, PIV-2000, L.C. by PB97-0893, A. Ac\'{\i}n and E.J. by a grant from MEC.
Financial support from the ESF is also acknowledged.

\end{document}